\documentclass[prl,twocolumn,tightenlines,superscriptaddress,a4paper,lengthcheck]{revtex4-1}
\usepackage{amsfonts}
\usepackage{amsmath}
\usepackage{amssymb}
\usepackage{graphicx}
\usepackage{graphicx,amssymb,amsbsy,color}
\usepackage{bm}
\usepackage[dvipdfm,colorlinks=true,linkcolor=blue,citecolor=blue,urlcolor=blue]{hyperref}

\begin{document}

\title{``Spin orders" in the supersolid phases in binary Rydberg-dressed Bose-Einstein condensates}
\author{C.-H. Hsueh}
\affiliation{Department of Physics, National Taiwan Normal University, Taipei 11677, Taiwan}
\author{Y.-C. Tsai}
\affiliation{Department of Physics, National Taiwan Normal University, Taipei 11677, Taiwan}
\affiliation{Department of Physics, National Changhua University of Education, Changhua
50058, Taiwan}
\author{K.-S. Wu}
\affiliation{Institute of Atomic and Molecular Sciences, Academia Sinica, Taipei 10617, Taiwan}
\author{M.-S. Chang}
\affiliation{Institute of Atomic and Molecular Sciences, Academia Sinica, Taipei 10617, Taiwan}
\author{W. C. Wu}
\affiliation{Department of Physics, National Taiwan Normal University, Taipei 11677, Taiwan}

\date{\today}

\begin{abstract}
We show that the five possible ordered states in a quantum spin-1/2 system with long-range exchange
interactions: N\'{e}el, ladder, Peierls, coincidence, and domain states,
can be realized in a binary Rydberg-dressed BEC system in the supersolid phase.
In such a system, blockade phenomenon is shown to also occur for pairs of different
excited-state atoms, which results in similar intra- and inter-species long-range interactions between
ground-state atoms. It suggests that a pseudo spin-1/2 system can be possibly formed
in the ground state of ultracold rudibium.
\end{abstract}
\maketitle

Supersolid (SS) is a matter state which simultaneously possesses crystalline
and superfluid (SF) properties. For the past four decades, SS states have
been studied by a number of authors\cite{Andreev,Chester,Leggett} and in
2004 it was first claimed that the SS state has been observed in a $^{4}$He
system\cite{Kim}, although in a later paper\cite{DKim} some inconsistency was
reported. An alternative and excellent candidate in which supersolidity can
be possibly observed is the atomic Bose-Einstein condensate (BEC) which
is clean and experimentally easily controllable
\cite{supersolid1,supersolid2,supersolid3,supersolid4,our supersolid}. In
particular, Rydberg dressed BEC poses a great potential to exhibit
the SS state that is intimately related to their long-range interaction.
Very recently it has been theoretically shown that SF-SS transition
in the Rydberg dressed BEC system is of the first order\cite{sf-ss}.

On the other hand, two- or multi-component system is of its own merit for
fundamental interest especially in regards to the role of the inter-species
interaction. For example, two-component BEC systems have been investigated
on various properties by a number of groups\cite{two1,two2,two3,two4,two5,two51,two6}.
Since the first experiment of two
coexisting condensates of two different hyperfine states of $^{87}$Rb was
realized in 1997\cite{two-com exp} and as the Rydberg dressing technique is mature,
it becomes possible to explore a SS state with internal degrees of freedom
(SS with a basis or a SS superlattice).
Without the long-range interaction, there are basically two phases: miscible
and insoluble, in a two-component SF system. With the long-range interactions, it is
interesting to see how the SS states behave, which should provide a
unique opportunity to study the complex phases of the two interpenetrating
quantum crystals.

This Letter attempts to investigate the ground-state phase diagram of a
binary Rydberg-dressed BEC system in the SS regime.
It will be shown that five distinct phases of the SS structures
are identified to exist in such a system.
Of particular interest, these ground-state phases are in a
strong analogy with the five possible spin orders in a quasi-two-dimensional quantum
spin-1/2 system with long-range exchange interactions.
Thus it suggests that a pseudo spin-1/2 system can be possibly formed
in the ground state of ultracold $^{87}$Rb.

\begin{figure}[t]
\begin{center}
\includegraphics[height=1.8in,width=3.5in]{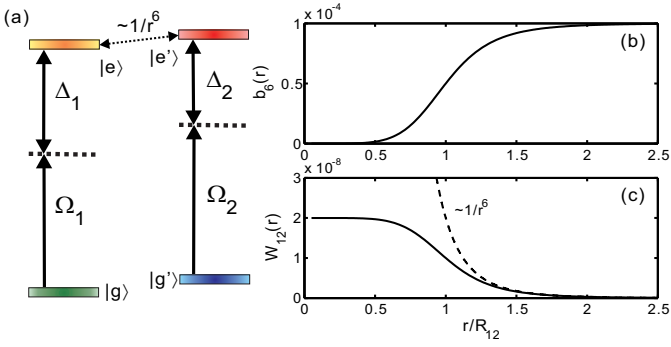}
\end{center}
\caption{(a) Schematic plot of a binary Rydberg-dressed BEC system
with the emphasis of the additional vdW interaction between different-species excited atoms.
Frame (b) shows the blockade phenomenon between different excited atoms by plotting the
coefficient $b_{6}$ vs. the distance $r$ (in units of $R_{12}$). Frame (c) shows the
spatial dependence of the long-range interaction between different ground-state atoms
by plotting the function $W_{12}$ vs. $r$, in units of $\hbar\Delta_{1}$ \cite{supersolid1}. }
\label{fig1}
\end{figure}

Fig.~\ref{fig1}(a) shows schematically a binary Rydberg-dressed BEC system
formed by two two-photon mechanisms. The binary system can be comprised of
two different (hyperfine) ground-state atoms excited to high-level Rydberg states.
Alternatively, the system can also be possibly formed by exciting the same
ground-state atoms to two different Rydberg states.
Under the rotating-wave approximation, the interaction Hamiltonian of
the system can be written as
\begin{eqnarray}
\hat{H} &=&\sum_{i<j}V^{\left( 11\right) }\left( r_{ij}\right) \hat{\sigma}%
_{ee}^{\left( i\right) }\hat{\sigma}_{ee}^{\left( j\right) }-\hbar \Delta
_{1}\sum_{i}\hat{\sigma}_{ee}^{\left( i\right) }  \notag \\
&+&\sum_{k<l}V^{\left( 22\right) }\left( r_{kl}\right) \hat{\sigma}%
_{e^{\prime }e^{\prime }}^{\left( k\right) }\hat{\sigma}_{e^{\prime
}e^{\prime }}^{\left( l\right) }-\hbar \Delta _{2}\sum_{k}\hat{\sigma}%
_{e^{\prime }e^{\prime }}^{\left( k\right) }  \notag \\
&+&\sum_{i,k}V^{\left( 12\right) }\left( r_{ik}\right) \hat{\sigma}%
_{ee}^{\left( i\right) }\hat{\sigma}_{e^{\prime }e^{\prime }}^{\left(
k\right) }+\hat{H}_{1}+\hat{H}_{2},  \label{HI}
\end{eqnarray}%
where $i,j,k,l$ are particle indices and $\hat{\sigma}_{\alpha \beta
}^{\left( i\right) }\equiv \left\vert \alpha _{i}\right\rangle \left\langle
\beta _{i}\right\vert $ define the corresponding transition and projection
operators with $\alpha $, $\beta =e$, $g$, $e^{\prime }$, $g^{\prime }$
denoting the excited and ground states of the first and the second species.
$V^{(pq)}(r)=C_{6}^{(pq)}/r^{6}$ ($p,q=1,2$) are, at a distance $r$,
the van der Waals (vdW) interactions between two identical/diverse Rydberg atoms
and $\Delta _{1}$, $\Delta _{2}$ are the laser detunnings.
$\hat{H}_{1}\equiv \left( \hbar \Omega _{1}/2\right) \sum_{i}\left[\hat{\sigma}
_{ge}^{\left( i\right) }+\hat{\sigma}_{eg}^{\left( i\right) }\right] $ and $%
\hat{H}_{2}\equiv \left( \hbar \Omega _{2}/2\right) \sum_{k}\left[ \hat{%
\sigma}_{g^{\prime }e^{\prime }}^{\left( k\right) }+\hat{\sigma}_{e^{\prime}
g^{\prime }}^{\left( k\right) }\right] $ with $\Omega_{1}$, $\Omega_{2}$ the Rabi
frequencies, are to be treated as perturbations for
$\Omega_{1}\ll \Delta_{1}$ and  $\Omega_{2}\ll\Delta_{2}$ of interest.
The interactions between ground-state atoms as well as between
Rydberg- and ground-state atoms are ignored as they are not important for
the studies. The effect of the foregoing interactions can be described in
terms of a $s$-wave scattering pseudopotential\cite{supersolid1}.

The interaction potential between two identical/diverse dressed atoms can be
well approximated by the Born-Oppenheimer (BO) energy surface $W_{G}$,
which can be determined from a many-body
perturbation expansion. Up to the fourth order involving couplings only to
the singly and doubly excited many-body states: $\left\vert GG^{\prime
}\right\rangle = \left\vert G\right\rangle \otimes \left\vert G^{\prime
}\right\rangle $, $\left\vert E_{i}G^{\prime }\right\rangle =
\left\vert E_{i}\right\rangle \otimes \left\vert G^{\prime }\right\rangle $,
$\left\vert E_{ij}G^{\prime }\right\rangle = \left\vert
E_{ij}\right\rangle \otimes \left\vert G^{\prime }\right\rangle $, $%
\left\vert GE_{i}^{\prime }\right\rangle = \left\vert G\right\rangle
\otimes \left\vert E_{i}^{\prime }\right\rangle $, $\left\vert
GE_{ij}^{\prime }\right\rangle = \left\vert G\right\rangle \otimes
\left\vert E_{ij}^{\prime }\right\rangle $, and $\left\vert
E_{i}E_{j}^{\prime }\right\rangle = \left\vert E_{i}\right\rangle
\otimes \left\vert E_{j}^{\prime }\right\rangle $, where $\left\vert
G\right\rangle \equiv\otimes _{k}\left\vert g_{k}\right\rangle $, $\left\vert
E_{i}\right\rangle \equiv\left\vert e_{i}\right\rangle \otimes _{k\neq
i}\left\vert g_{k}\right\rangle $, $\left\vert E_{ij}\right\rangle
\equiv\left\vert e_{i}e_{j}\right\rangle \otimes _{k\neq i,j}\left\vert
g_{k}\right\rangle $, $\left\vert G^{\prime }\right\rangle \equiv\otimes
_{k}\left\vert g_{k}^{\prime }\right\rangle $, $\left\vert E_{i}^{\prime
}\right\rangle \equiv\left\vert e_{i}^{\prime }\right\rangle \otimes _{k\neq
i}\left\vert g_{k}^{\prime }\right\rangle $, and $\left\vert E_{ij}^{\prime
}\right\rangle \equiv \left\vert e_{i}^{\prime }e_{j}^{\prime }\right\rangle
\otimes _{k\neq i,j}\left\vert g_{k}^{\prime }\right\rangle$, we obtain
\begin{eqnarray}
W_{G}&=&\sum_{i<j}W_{11}\left( r_{ij}\right) +\sum_{k<l}W_{22}\left(
r_{kl}\right) +\sum_{i,k}W_{12}\left( r_{ik}\right)\nonumber\\
&+&\text{constant,}
\label{WG}
\end{eqnarray}
where ($p,q=1,2$) \cite{supersolid1}
\begin{eqnarray}
W_{pq}(r)={\tilde{C}_{6}^{(pq)}\over R_{pq}^{6}+r^{6}}.
\label{Wpq}
\end{eqnarray}
Here $\tilde{C}_{6}^{\left(pq\right) }=\nu _{p}\nu _{q}C_{6}^{\left( pq\right) }$ and
$R_{pq}=[C_{6}^{\left( pq\right) }/\left( 2\hbar \Delta _{pq}\right)]^{1/6}$ with
$\nu _{l}\equiv \left( \Omega _{l}/2\Delta _{l}\right) ^{2}$ ($l=1,2$),
$\Delta_{11}\equiv \Delta_1$, $\Delta_{22}\equiv \Delta_2$, and
$\Delta _{12}\equiv(\Delta _{1}+\Delta _{2})/2$ are
the intra- and inter-species effective coupling constant and blockade radius, respectively.
In the weak Rabi frequency limit of interest, $\nu_{1},\nu_{2}\ll 1$,
the whole dressed state can be expressed as a direct
product of two individual dressed states:%
\begin{widetext}
\begin{eqnarray}
\left\vert \Psi\right\rangle  &=&\left\vert \Psi_{1}\right\rangle\otimes
\left\vert \Psi_{2}\right\rangle\approx b_{1}\left\vert gg\right\rangle \otimes\left\vert g^{\prime
}g^{\prime}\right\rangle +{b_{2}\over \sqrt{2}}\left(  \left\vert
ge\right\rangle +\left\vert eg\right\rangle \right)  \otimes\left\vert
g^{\prime}g^{\prime}\right\rangle
+b_{3}\left\vert ee\right\rangle \otimes\left\vert g^{\prime}g^{\prime
}\right\rangle \nonumber\\
&+&\frac{b_{4}}{\sqrt{2}}\left\vert gg\right\rangle
\otimes\left(  \left\vert g^{\prime}e^{\prime}\right\rangle +\left\vert
e^{\prime}g^{\prime}\right\rangle \right)
+b_{5}\left\vert gg\right\rangle \otimes\left\vert e^{\prime}e^{\prime
}\right\rangle +\frac{b_{6}}{2}\left(  \left\vert ge\right\rangle +\left\vert
eg\right\rangle \right)  \otimes\left(  \left\vert g^{\prime}e^{\prime
}\right\rangle +\left\vert e^{\prime}g^{\prime}\right\rangle \right),
\label{dressed state}
\end{eqnarray}
\end{widetext}
where the coefficients $b_{1}$ to $b_{6}$ are of
order $1$, $\sqrt{\nu _{1}}$, $\sqrt{\nu _{2}}$, $\nu _{1}$, $\nu _{2}$, and
$\sqrt{\nu _{1}\nu _{2}}$, respectively. Fig.~\ref{fig1}(b) plots $b_{6}$ as a
function of the distance $r$. It indicates clearly that the blockade effect also exists
between the diverse excited Rydberg atoms. This in turn leads to
similar intra- and inter-species long-range interactions between
ground-state atoms, as shown explicitly by the plot
of the inter-species potential $W_{12}$ in Fig.~\ref{fig1}(c).

As this paper concerns about the SS ground-states not too close
to the SF-SS transition, mean-field Gross-Pitaevskii (GP) treatment should be
sufficient for the study. In the dimensionless scheme, the GP energy functional
of the system can be written as
\begin{equation}
E=E_{0}+E_{\mathrm{int}},
\label{E}
\end{equation}%
where the single-particle part
\begin{equation}
E_{0}=\sum_{i=1,2}\int d{\vec \rho}\left[ \left\vert \nabla \psi
_{i}({\vec \rho})\right\vert ^{2}/2+\rho ^{2}\left\vert \psi _{i}({\vec \rho})
\right\vert ^{2}/2\right]
\label{E0}
\end{equation}
and the interaction part
\begin{equation}
E_{\mathrm{int}}=\frac{1}{2}\sum_{i,j=1,2}\int d{\vec \rho}~d{\vec {\rho^\prime}}
V_{ij}(\bar{\rho})|\psi _{i}({\vec \rho})|^{2}|\psi _{j}({\vec {\rho^\prime}})|^{2}
\label{Eint}
\end{equation}
with $\bar{\rho}\equiv |{\vec \rho}-{\vec {\rho^\prime}}|$. In (\ref{Eint}), the intra- and
inter-species interactions comprised of contact and long-range parts are
isotropic and
\begin{equation}
V_{ij}(\bar{\rho})=\gamma _{ij}\delta (\bar{\rho})+U_{ij}(\bar{\rho})=
\gamma_{ij}\delta \left( \bar{\rho}\right)+\frac{\alpha _{ij}}{r_{ij}^{6}+\bar{\rho}^{6}},
\label{V}
\end{equation}
where $\alpha _{ij}=m\tilde{C}_{6}^{\left( ij\right) }/\left( \hbar
^{2}l_{h}^{4}\right) $ are the strengths of the long-range interactions, $%
r_{ij}=R_{ij}/l_{h}$ are the blockade radii, and $\gamma _{ij}=4\pi a_{ij}$
are the two-dimensional coupling constants with $a_{ij}$ the effectively
$s$-wave scattering lengths. In Eqs.~(\ref{E0}) and (\ref{Eint}),
it is assumed that the two-species Rydberg-dressed BECs are trapped respectively
in the harmonic potentials $U_{\mathrm{trap}}^{\left( i\right) }
=m_{i}\omega _{i}^{2}\left(\rho ^{2}+\lambda _{i}^{2}z^{2}\right) /2$ ($i=1,2$)
in the cylindrical coordinates ${\bf r}=\left( \rho ,\phi ,z\right) $ with
$\omega _{i}$ and $\lambda _{i}$ being the radius frequencies and the aspect ratios, respectively.
The length and energy scales are thus $l_{h}\equiv \sqrt{\hbar /m\omega }$ and $\hbar \omega$ respectively.
Appropriate for the SS regime ($\lambda _{i}\gg 1$), the calculation is to
be treated quasi-two-dimensionally and in this limit, the population
is determined by $\int \left\vert \psi _{i}\right\vert ^{2}dxdy=N$\cite{qusi-2d}.
In actual calculations, for simplicity, it is assumed that $m_{2}=m_{1}\equiv m$, $\omega _{2}=\omega
_{1}\equiv\omega$, and $\lambda _{2}=\lambda _{1}\equiv \lambda$.
Moreover, we also assume that all the
blockade radii $r_{ij}$ are the same, labeled as $r_{c}$.

\begin{widetext}

\begin{figure}[t]
\begin{center}
\includegraphics[height=2.4in,width=4.6in]{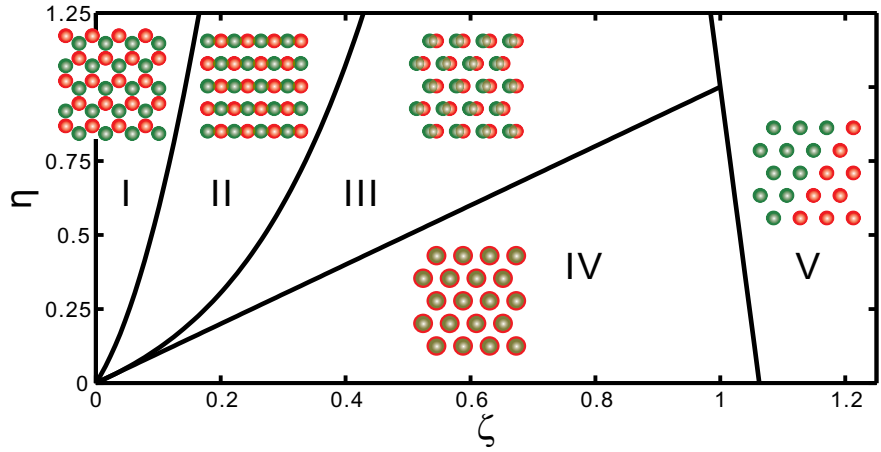}
\end{center}
\vspace{-0.5cm}
\caption{Phase diagram of a binary Rydberg-dressed BEC system showing five distinct phases (I--V)
in the SS regime. Considering  green droplet to be a pseudospin up ($\uparrow$) and
red droplet to be a pseudospin down ($\downarrow$), phase I--V correspond to N\'{e}el, ladder,
Peierls, coincidence, and domain states in the quantum spin-1/2 system.
($\zeta=1,\eta=1$) is a tricritical point.}
\label{fig2}
\end{figure}

\end{widetext}

\begin{widetext}

\begin{figure}[t]
\includegraphics[trim=0.000000in 0.000000in -0.004502in 0.000000in,
height=2.1037in,width=6.6396in]{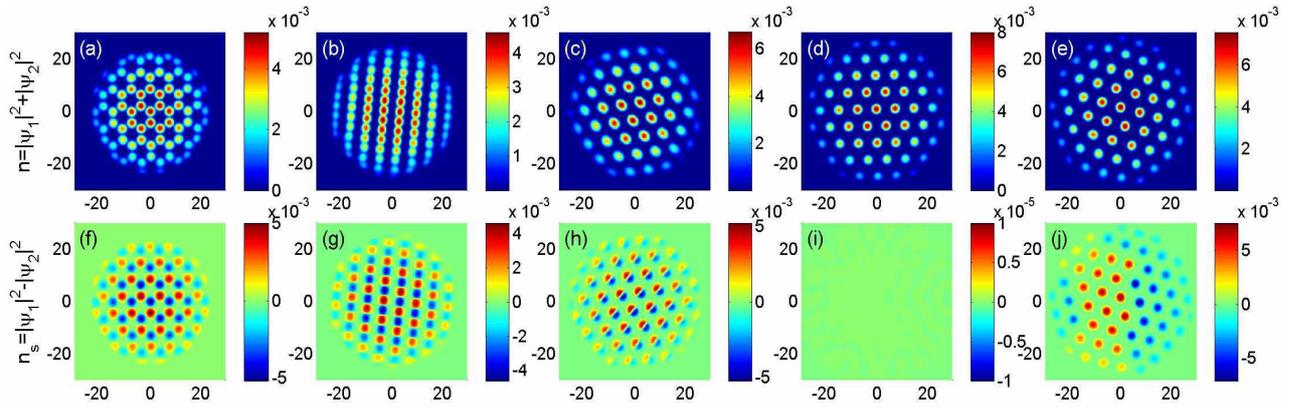}
\caption{Charge (top row) and spin (bottom row) orders
of the five phases associated with a fixed $\eta=0.75$
and $\zeta=0.03125$ (a,f), $0.25$ (b,g), $0.5$ (c,h), $0.875$ (d,i), and
$1.25$ (e,j) respectively. Other parameters used are $N\gamma_{11}=N\gamma_{22}=10^4$, $r_c=5$,
and $N\alpha_{11}=N\alpha_{22}=6.25\times 10^7$.}
\label{fig3}
\end{figure}

\end{widetext}

Fig.~\ref{fig2} shows the phase diagram of the system which is characterized by
two parameters $\eta\equiv a_{12}/a_{11}$ and
$\zeta \equiv \alpha _{12}/\alpha _{11}$ associated with the ratios of
the short-range (contact) and long-range interactions respectively.
Due to the robustness of the two hexagonal
sublattices, five stable distinct phases (I)--((V) are unambiguously identified.
In our calculations, the parameter space is first discretized into 40$\times$20 points
in the $\zeta$-$\eta$ plane. The phase boundaries are then carefully reexamined and demarcated
between two neighboring phases.
In terms of density structures (orders), they correspond to
(I) honeycomb lattice (one droplet surrounded by three diverse ones), (II) rectangle lattice
(one droplet surrounded by four diverse ones), (III) triangular lattice with
dimeric basis, (IV) triangular lattice without basis, and (V)
the domain structures. Demarcated by the $\eta =1$ line,
there are four and five phases above and below it, respectively.
An explicit example of how density profiles behave
in each phase is given in the top row of Fig.~\ref{fig3}.

For the current two-species system, it is interesting and useful to introduce a
two-component spinor wavefunctions, $\Psi \equiv \left(
\psi _{1},\psi _{2}\right) ^{t}$. In terms of the ``charge"
$n\equiv \Psi ^{\dag }\sigma_0\Psi =|\Psi _{1}|^{2}+|\Psi _{2}|^{2}$
and the ``spin"
$n_s\equiv \Psi ^{\dag }\sigma _{3}\Psi =|\Psi _{1}|^{2}-|\Psi _{2}|^{2}$,
the interaction energy functional in (\ref{Eint}) can be rewritten as
\begin{eqnarray}
E_{\mathrm{int}} &=&\frac{1}{4}\int d{\vec {\rho}}~d{\vec {\rho^\prime}}
\left[ V_{11}(\bar{\rho})-V_{22}(\bar{\rho})\right] n\left({\vec{\rho^\prime}}\right)
n_s\left({\vec \rho }\right)   \notag \\
&+&\frac{1}{2}\int d{\vec {\rho}}~d{\vec {\rho^\prime}}
J_s(\bar{\rho})n_s\left({\vec{\rho^\prime}}\right)
n_s\left({\vec \rho }\right)\nonumber\\
&+&\frac{1}{2}\int d{\vec {\rho}}~d{\vec {\rho^\prime}}
J_{\mathrm{n}}(\bar{\rho})n\left({\vec{\rho^\prime}}\right)
n\left({\vec \rho }\right),
\label{eint}
\end{eqnarray}%
where
\begin{eqnarray}
J_n(\bar{\rho}),J_s(\bar{\rho})
=\frac{1}{4}\left[ V_{11}(\bar{\rho})+V_{22}(\bar{\rho})\pm 2V_{12}(\bar{\rho})\right].
\label{hds}
\end{eqnarray}
The first term of $E_{\mathrm{int}}$ corresponds to a coupling between
charge $n$ and spin $n_s$, while the second and third terms correspond to spin and
charge channels respectively. In the case $V_{11}\approx V_{22}$, the first term
is negligible and thus the second and third terms dominate.
In view of the second and third lines in (\ref{eint}), the spin-channel term
has a strong analogy to the one for a spin liquid to which $J_s(\bar{\rho})$ is considered to
be the corresponding {\em exchange} coupling. Whereas for the charge channel,
$J_n(\bar{\rho})$ is considered to be the corresponding {\em Coulomb} interaction.

By separating out the contact and the long-range parts of the interaction
and under the approximation $U_{11}=U_{22}$ and $U_{12}=\zeta U_{11}$ for the long-range part,
the couplings in (\ref{hds}) can be rewritten as
\begin{eqnarray}
J_n(\bar{\rho}) ,J_s(\bar{\rho}) &=&{1\over 2}\left[
\gamma _{11}\left( 1\pm \eta \right) \delta (\bar{\rho})+\left( 1\pm \zeta \right)
U_{11}(\bar{\rho})\right].\nonumber\\
\label{Vns}
\end{eqnarray}
In view of (\ref{Vns}) and considering the long-range effect in the charge channel,
the coefficient $1+\zeta$ is always
positive and it is the main cause for the robustness of the crystallization structure of the
system\cite{crystal1,crystal2,crystal3}.
More precisely, the robustness of the hexagonal sublattice for each species is
due to the strong isotropic intra-species long-range interactions $U_{11}$ and $U_{22}$.
For the spin channel, in contrast, the coefficient $1-\zeta$ is strongly magnitude
and sign dependent. Consequently how the sublattices correlate with each other will depend
strongly on the value of $\zeta$, or more precisely on the
inter-species long-range interaction $U_{12}=\zeta U_{11}$.

The phase diagram in Fig.~\ref{fig2} can thus be alternatively interpreted in terms of
spin orders (or magnetization) treating the system of
pseudospin-1/2 \cite{blundell}.
With a relatively large and positive exchange coupling $\zeta^\prime\equiv 1-\zeta$,
phase I corresponds to the N\'{e}el state on a honeycomb lattice.
Phase II is so-called the $n$-leg spin-ladder state by relaxing the coupling $\zeta^\prime$.
Phase III is the spin-Peierls state by further decreasing
the coupling $\zeta^\prime$. Phase IV is the
spin-up/spin-down coincidence state with relatively small but positive $\zeta^\prime$.
Phase V is the domain state due to the sign change of the exchange coupling
($\zeta^\prime<0 $). Moreover, based on the long-range exchange coupling given
in Eq.~(\ref{Vns}), it is comprehended that when $\zeta^\prime$ is positive,
antiferromagnetism (AFM) dominates and as a matter of fact, the larger $\zeta$ or
the smaller $\zeta^\prime$ is, the more overlap between the two sublattices is.
However, when $\zeta^\prime$ changes sign to be negative,
ferromagnetic (FM) dominates and consequently domains form.
Therefore, one can classify phase I--IV belonging to AFM, while
phase V belonging to FM. An explicit example of how the spin order behaves
in each phase is given in the bottom row of Fig.~\ref{fig3}.

Fig.~\ref{fig3} gives explicit examples of charge (top row) and spin
(bottom row) orders associated with various phases in
Fig.~\ref{fig2}. A fixed $\eta=0.75$ is used and $\zeta$ are varied to be
$0.03125$, $0.25$, $0.5$, $0.875$, and $1.25$ respectively for
frame (a)\&(f), (b)\&(g), (c)\&(h), (d)\&(i), and (e)\&(j), corresponding to phase I to V.
Other parameters used are $N\gamma_{11}=N\gamma_{22}=10^4$, $r_c=5$,
and $N\alpha_{11}=N\alpha_{22}=6.25\times 10^7$. Justifications of these parameters can be
referred to Ref.~\cite{rotating} that warrants the validity of the
blockade regime and the use of the effective long-range interactions.

By connecting the neighboring phases for a small and same $\zeta$ and by varying $\eta$,
one is able to understand the role of the contact interactions. As a matter of fact,
the increase of $\eta$ (or the ratio of $\gamma_{12}/\gamma_{11}$) will increase
the separation of the nearby diverse droplets, resulting in different lattices.

In summary, five distinct phases are identified to exhibit in the
two-species Rydberg-dressed BEC system in the supersolid regime. These phases are
in one-to-one correspondence to the five spin orders, N\'{e}el, ladder, Peierls, coincidence,
and domain states in a quantum spin-1/2 system with long-range exchange
interactions and suggest a possible pseudo spin-1/2 system forming in
the ground state of rudibium.

We are grateful to Prof. Ming-Che Chang for many useful comments on
this work. The supports from the National Science Council of Taiwan
(under the grant No. NSC 99-2112-M-003-006-MY3)
and the National Center of Theoretical Sciences of Taiwan are acknowledged.


\begin{thebibliography}{99}
\bibitem{qusi-2d} When the Rydberg-dressed system is strongly confined in $z$-axis
by a $z$-direction harmonic potential $(\omega \lambda)^{2}z^{2}/2$\ and relatively freely
moving in the $xy$\ plane, the 3D wavefunction can be
approximated by $\psi _{i}^{\mathrm{(3D)}}(\mathbf{r})=\psi _{i}^{\mathrm{%
(2D)}}({\vec \rho })\varphi _{g}(z)$ with $\varphi _{g}(z)=
\left( \omega \lambda /\pi \right) ^{1/4}\exp (-\omega \lambda z^{2}/2)$.
As the softened range $r_{ij}$ is much longer than the width of $\varphi
_{g}^{2}$, the 2D long-range interactions can be well approximated
by $U_{ij}^{\mathrm{(2D)}}({\vec \rho }-{\vec {\rho^\prime}})\approx
\alpha _{ij}/[r_{ij}^{6}+\bar{\rho}^{6}]\times \int \varphi
_{g}(z)^{2}\varphi _{g}(z^{\prime })^{2}dzdz^{\prime }=\alpha
_{ij}/[r_{ij}^{6}+\bar{\rho}^{6}]$.
\end{thebibliography}
\end{document}